\documentclass[a4paper]{revtex4}
\usepackage{graphicx}
\usepackage{fancyhdr}
\usepackage{amsmath}
\usepackage{color}

\begin{document}

\title{ Higgs pair production from color octet scalars and vectors}

%

\author{Tsedenbaljir Enkhbat}
\affiliation{Insitute of Physics and Technology, Mongolian Acadmey of Sciences, Ulaanbaatar 13330, MONGOLIA}

\begin{abstract}
Higgs pair production is studied in the extension of the Standard Model by color octet scalar and vector particles in TeV mass range for the LHC. The relevant parameters are their masses and portal couplings to the Higgs boson.
 The rate is enhanced considerably in some parts of the parameter space compared to the SM which are allowed by the latest measured bounds on the Higgs decay to $ZZ^*$. The model will be tested experimentally at the upcoming LHC run.
\end{abstract}

\maketitle

\thispagestyle{fancy}


\section{Introduction}
The discovery of the Higgs boson sets the stage for the next experimental quests including the nature and the origin of electroweak symmetry breaking~\cite{atlas:2012gk}.
The Higgs self coupling is the measure of the electroweak symmetry breaking (EWSB) in the Standard Model (SM) and can be probed 
experimentally by Higgs pair production processes~\cite{Eboli:1987dy,Jikia:1992mt}.  Their rate is, however, predicted to be quite small in the SM, making the experimental measurements very challenging.  For example, the Higgs pair production rate at the LHC is estimated to be 30$\sim$40~fb for the 14~TeV run, thus requires very high luminosity of 3~ab$^{-1}$ for 3~$\sigma$ evidence~\cite{Barger:2013jfa}. This is chiefly due to the cancellation between the triangle and box top quark loop diagrams. On the other hand, any new particle with a substantial coupling to the SM Higgs can easily offset this and the enhancement of the rate can  be sometimes up to few orders of magnitude~\cite{Belyaev:1999mx}. Any colored particle is interesting in their own right as it can be created at the LHC via strong interactions. For different motivations and in variety of contexts color octets and triplets have been studied extensively in the past. One of the most recurring motivation is that they may play a potentially important role in the evolution of the early Universe. Often times new fields, especially colored particles are needed for sufficiently strong electroweak phase transition, in the framework of electroweak baryogenesis.
	
	In this report we present a simple extension of the SM by two color octet fields, scalar and vector, which have sizable portal couplings to the Higgs doublet. Instead of assigning particular charges we take two cases: both are real and both are complex. 

\section{The effect of vector and scalar fields on Higgs pair production}
\begin{figure}[ht]
\centering
\includegraphics[width=80mm]{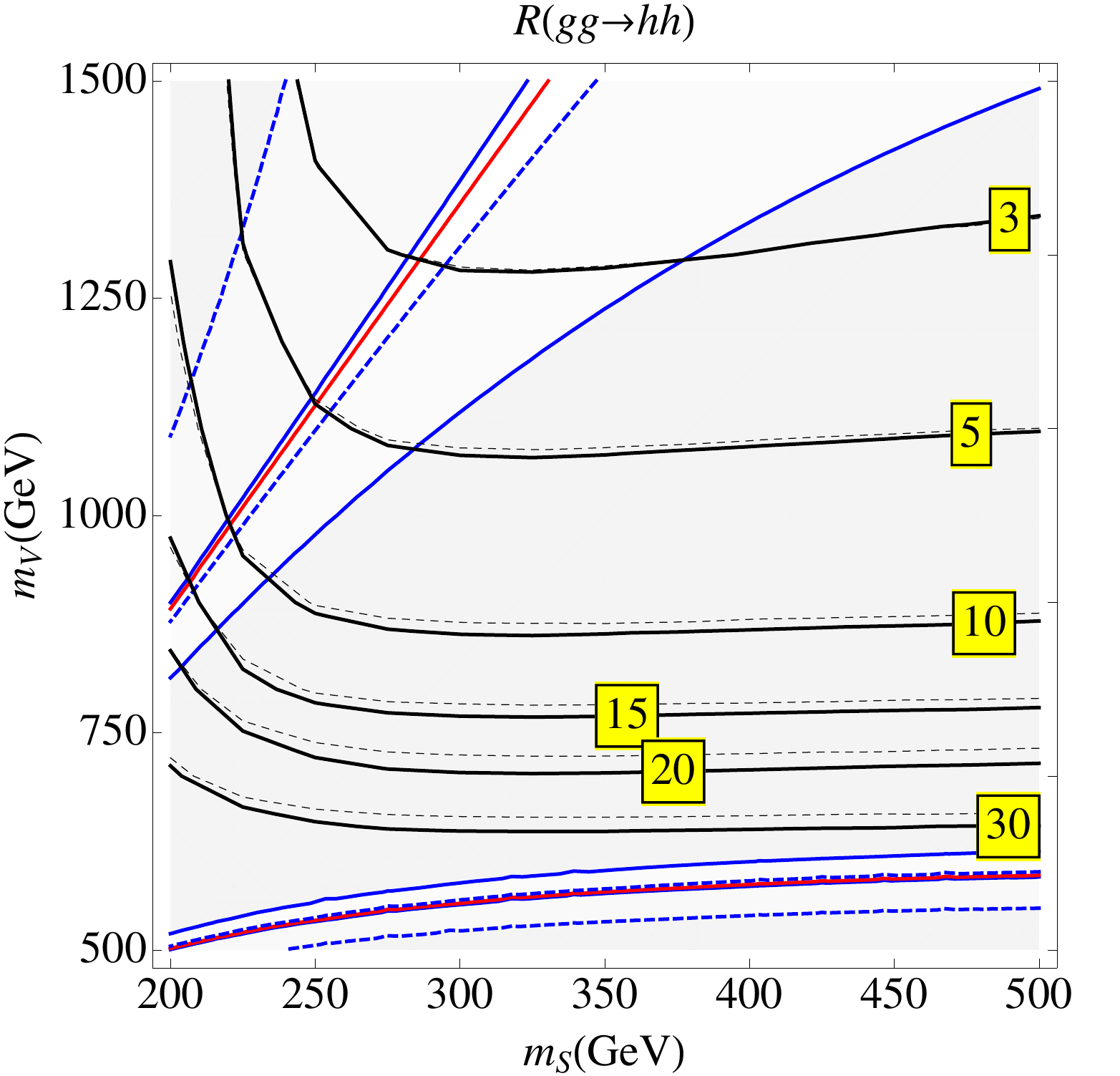}
\includegraphics[width=80mm]{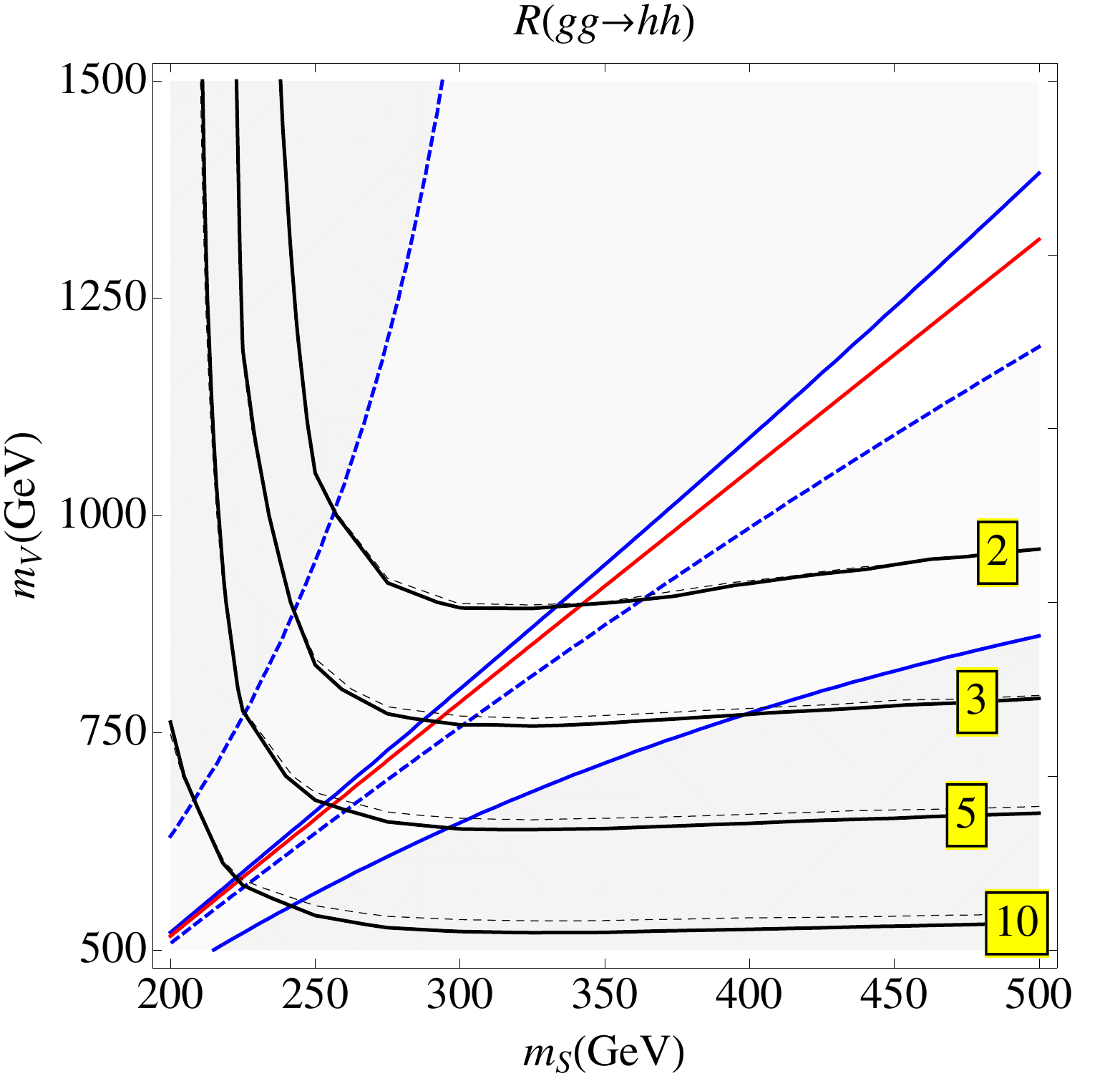}
\includegraphics[width=80mm]{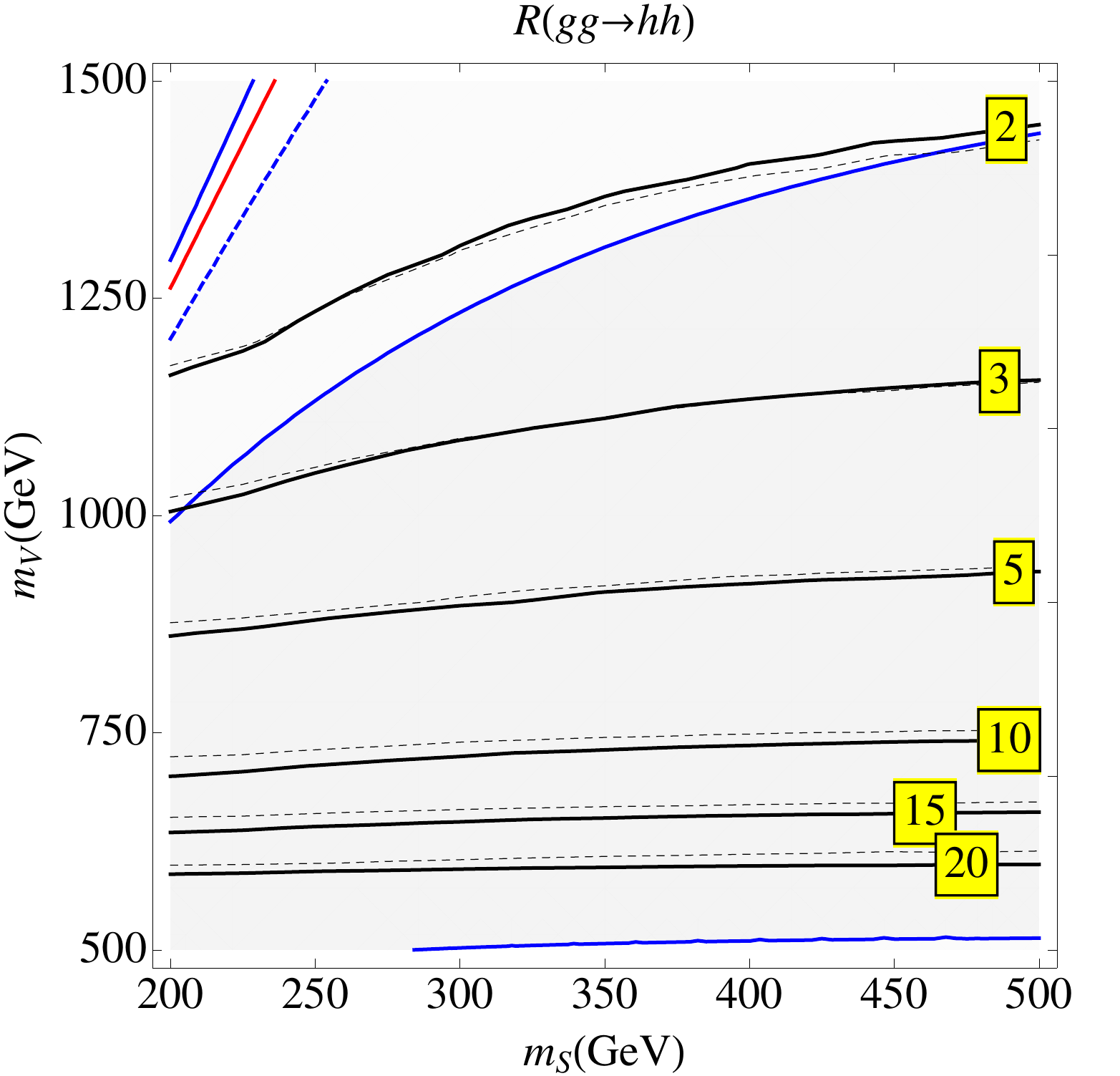}
\includegraphics[width=77mm]{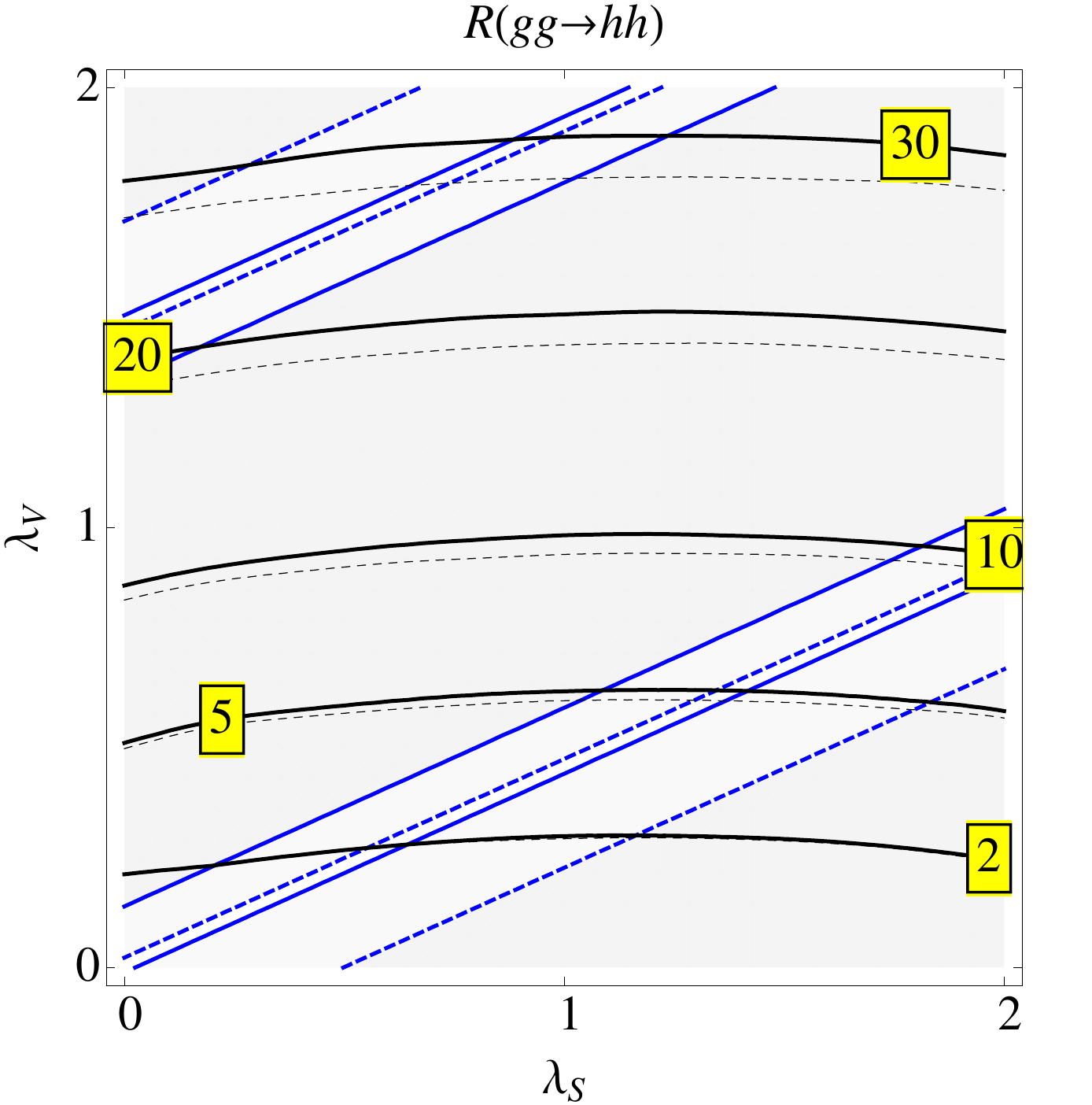}
\caption{The Higgs pair production rate compared to the SM in the presence of color octet, real scalar and vector particles. The first three are scan over the masses of these particles. The black solid ( dashed) lines are for the $\sqrt{s}=8$~TeV (13~TeV) run and their are labeled by the enhancement factors over the SM. The solid ( dashed) blue lines are the bounds from $h \to ZZ^*$ for CMS (ATLAS) and the shaded regions are excluded. The last graph is the scan over the couplings.} \label{figure1}
\end{figure}

\begin{figure}[ht]
\centering
\includegraphics[width=80mm]{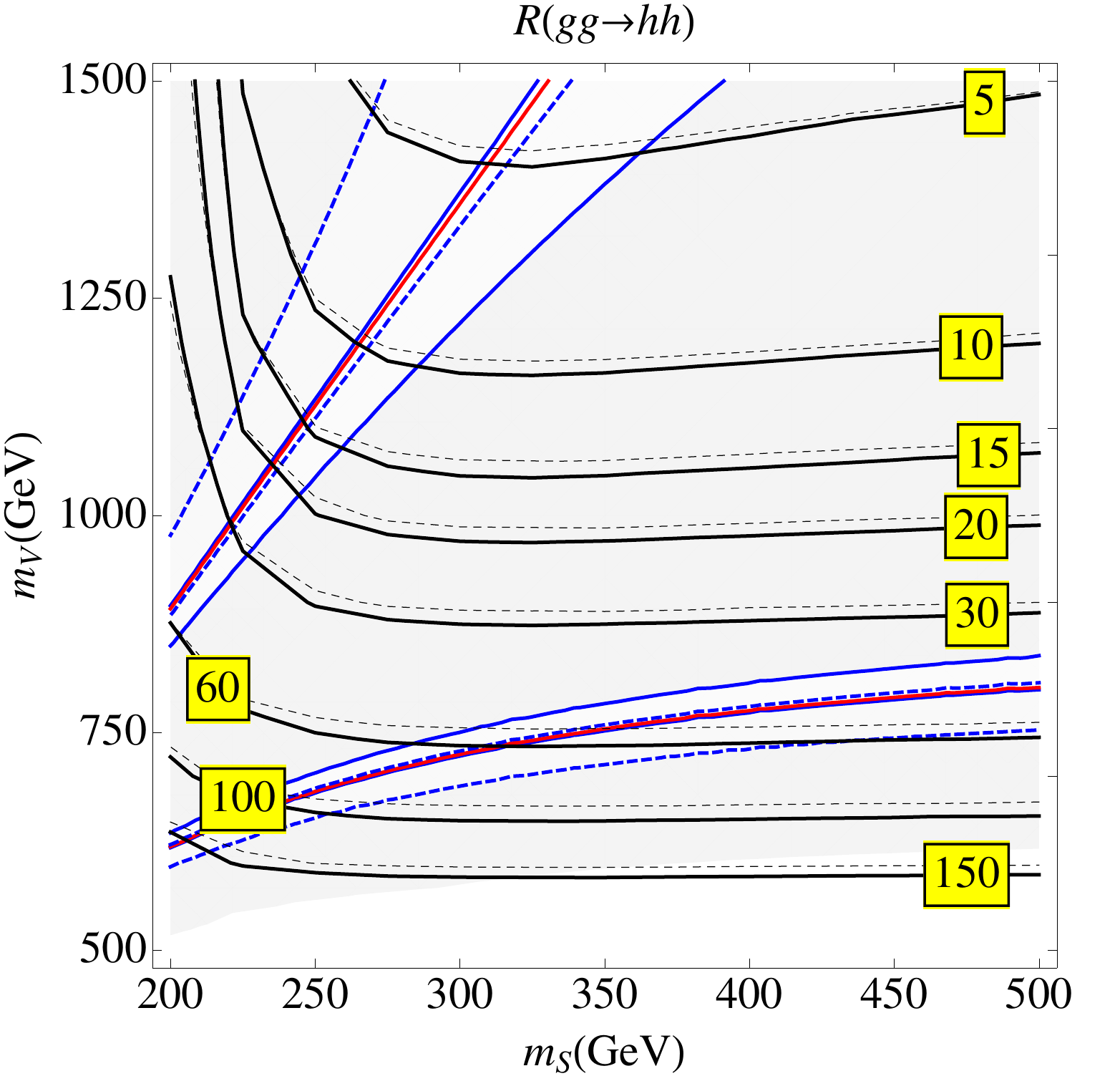}
\includegraphics[width=80mm]{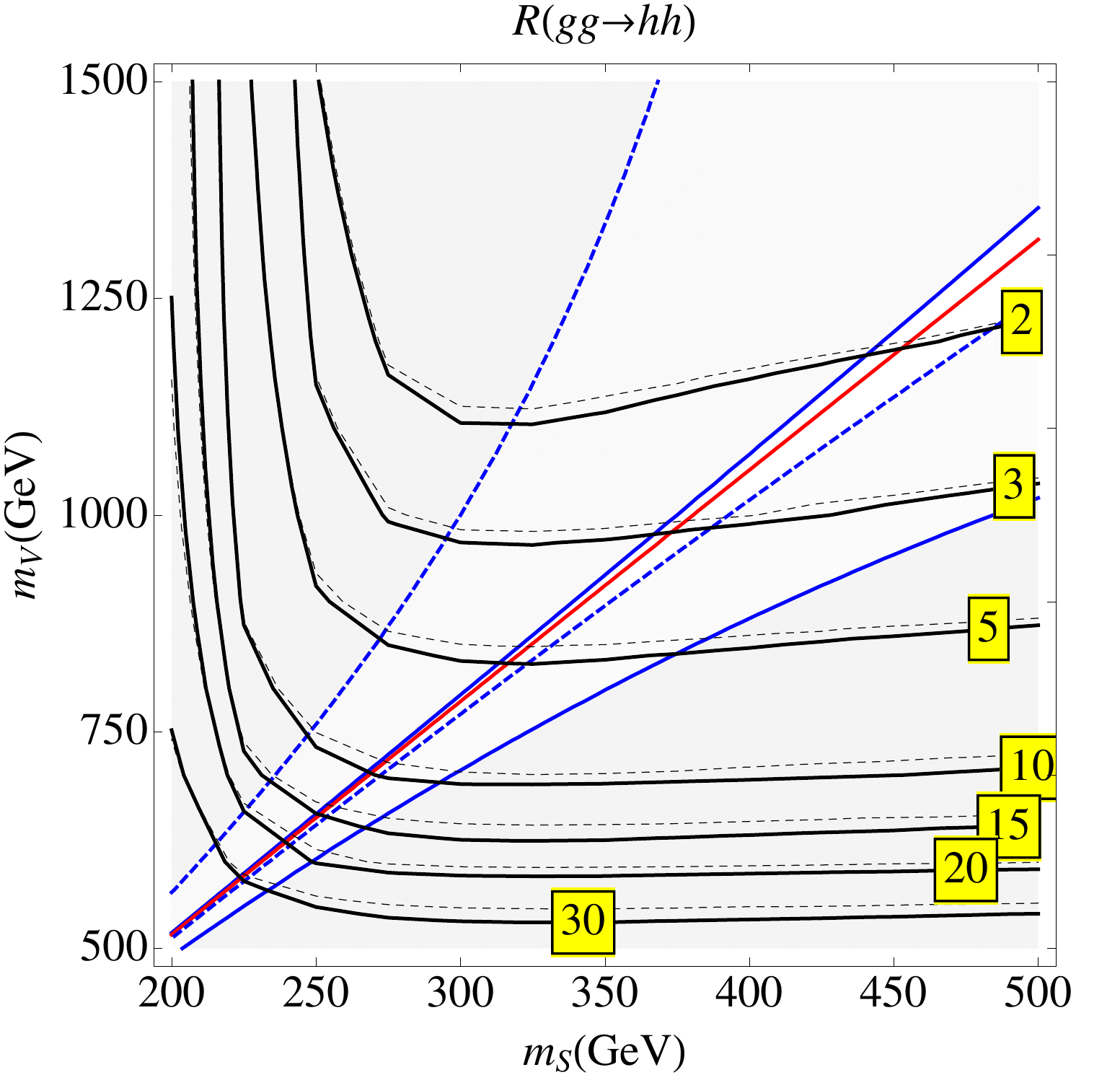}
\includegraphics[width=80mm]{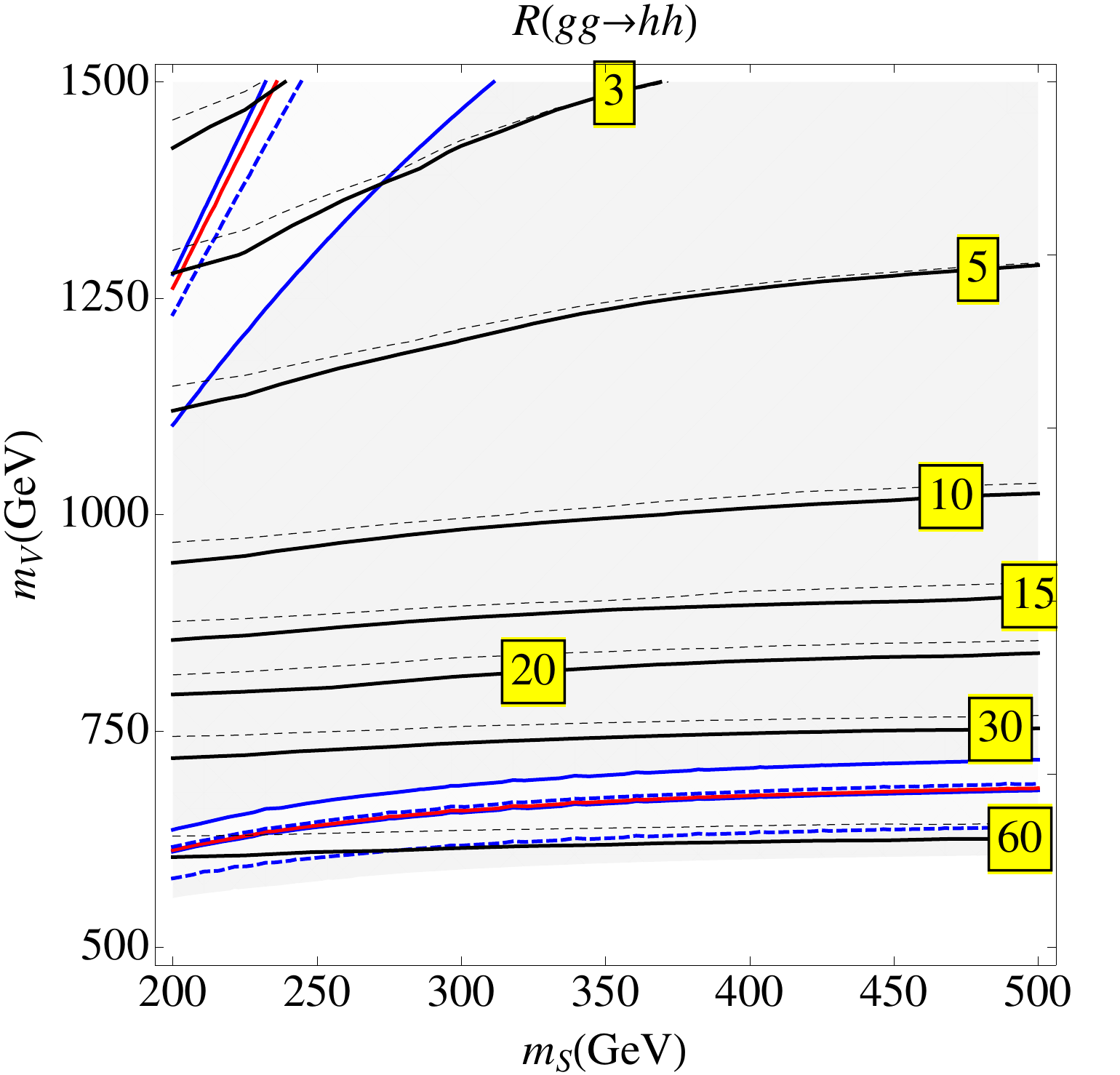}
\includegraphics[width=77mm]{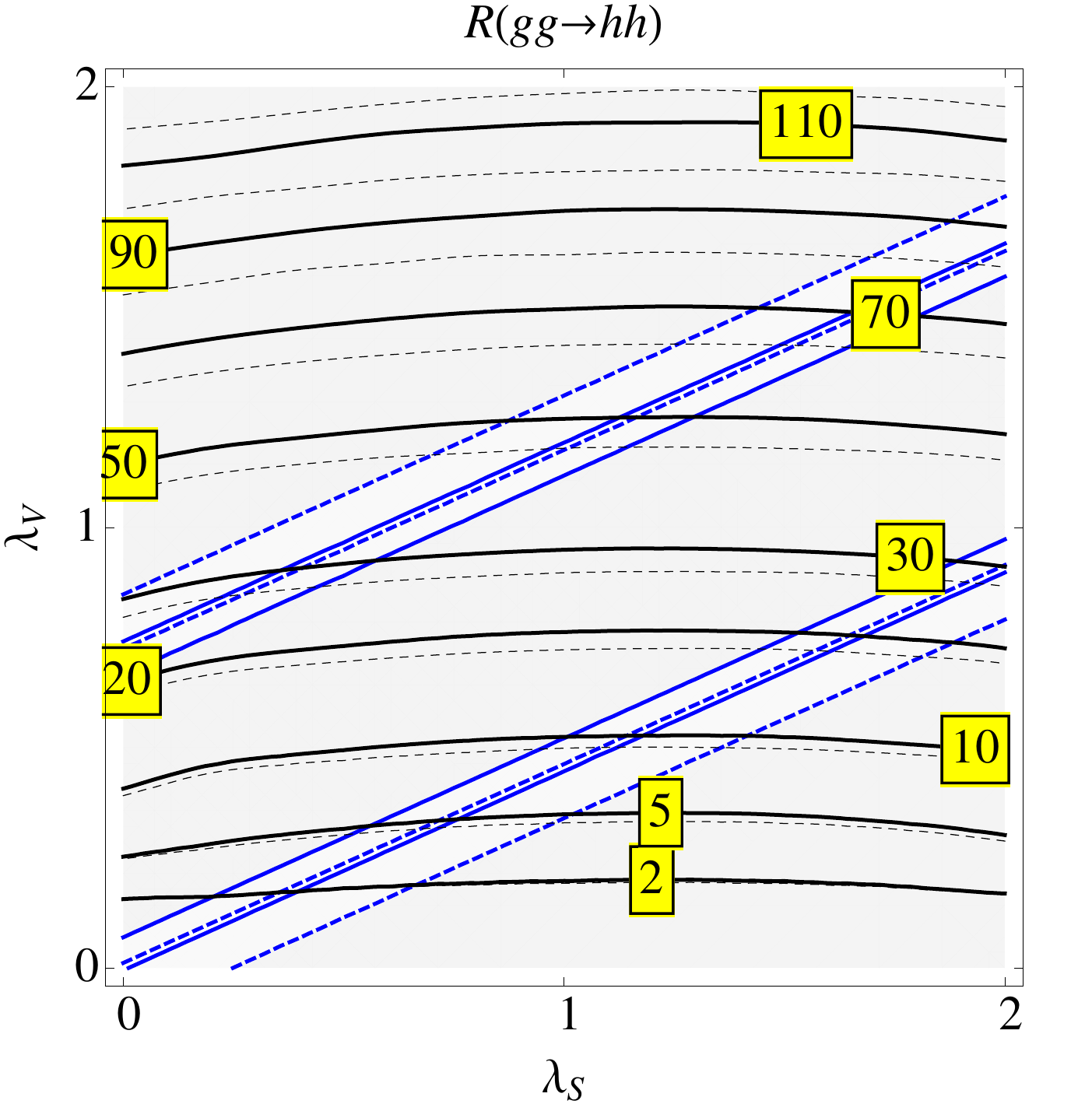}
\caption{The same as Fig.~\ref{figure1} in the case of  complex scalar and vector particles.} \label{figure2}
\end{figure}

Scalar colored particles with portal couplings to Higgs have been considered in a number of studies.
Their effect on Higgs pair production can be quite large especially for a negative portal couplings.  When there is only one such particle it is nearly impossible to have any appreciable effect while maintaining a good agreement with the experimentally allowed range for Higgs decay to diphoton and $ZZ^*$ channels. Now only possibility remains is to have a large corrections of roughly twice the SM contributions to diphoton amplitude but with opposite in sign. With low mass regions are being excluded by the accumulating data at the LHC one is forced to choose higher mass at the same time with stronger couplings. This quickly makes the vacuum unstable, therefore physically unacceptable. The situation is more relaxed if there are more than one colored particles as they open up more possibilities.
	In the present talk we consider a Lorentz vector and scalar particles in octet representation. For the sake of not making the graphs too cumbersome  we plot the latest constraint on such particles from $h \to ZZ^*$ channel from ATLAS and CMS. Depending on charge assignments $h\to \gamma \gamma $ can be accommodated. Due to space constraint we give two different cases here out of three we have presented at the symposium. These are both vector and scalar are either complex or  real. The leading production mechanism for the Higgs pair production at the LHC is the gluon--gluon fusion. There are two non interfering amplitudes belonging to the same and opposite helicity initial gluons. The first one receives contributions from both the triangle and box diagrams while the latter does only from the box diagrams. In the SM, the triangle and box diagrams exactly cancel at the threshold energy of two Higgs production. This balance is altered whenever there is a new particle. The relevant portal interactions are given as follows
\begin{align}
{\cal L}={\cal L}_{SM}+{\cal L}(S,V)_{kin}+{\cal L}(S,V)_{int}-{\lambda_s} |S|^2|H|^2+\lambda_v V_\mu V^{\mu*}|H|^2
\label{eq:Lag}
\end{align}	
for real octets and these terms should be replaced by $-\frac{\lambda_s}{2} S^2|H|^2+\frac{\lambda_v}{2} V_\mu V^{\mu}|H|^2$ for the complex case. The other terms are represent the SM part, the kinetic terms for the new particles and their interactions that do not include the SM particles respectively.
	The expressions for fermion and scalar contributions are well known and can be found elsewhere~\cite{Eboli:1987dy}. As for the vector contribution we take the following expression given in Ref~\cite{Jikia:1992mt}. $F_{tri}$, $F_{box}$ are the amplitudes for Here $m$ and $m_h$ are the masses of the vector particle in the loop and Higgs, $\lambda_v\,(\lambda_s)$ is the portal coupling of the vector (scalar) to Higgs. $C_v$ is the Dynkin index. It is 3 (6) for real (complex) color octet. The functions $C_{ab}$, $D_{abc}$ etc are the two and three point Passarino-Veltman loop functions. 
\begin{align}
F^V_{tri}&=\left(1+\frac{3m_h^2}{s-m_h^2}\right)\frac{C_v\lambda_vv^2}{m^2}\left(8sC_{ab}+(6m^2+m_h^2)\left(1+2m^2C_{ab}\right)\right),\\
F^V_{box}&=C_v\left(\frac{\lambda_vv^2}{m^2}\right)^2\frac{m^2}{2s}\left(4sm^2\left(D_{abc}+D_{bac}+D_{acb}\right)-4sC_{ab}\right.\\
&\left. + \frac{m_h^4-2m_h^2m^2+12m^4}{2sm^2}\left((t-m_h^2)m^2C_{ac}+(u-m_h^2)C_{bc}-(tu-m_h^4) D_{acb}\right)\right.\nonumber\\
&\left. -2sm^2\left(D_{abc}+D_{bac}+D_{acb}\right)\right)\nonumber,\\
G^V_{box}&=-C_v\left(\frac{\lambda_vv^2}{m^2}\right)^2\frac{m^2}{2s}\left(2(tu-m_h^4)\left(D_{abc}+D_{bac}+D_{acb}\right)-4sC_{ab}
\right.\\
&\left.+\frac{1}{2m^2(tu-m_h^4)}\left((4m^2(t-m_h^2)^2-M^4t)(2(t-m_h^2)C_{ac}+(t-m_h^2)^2D_{bac})
\right.\right.\nonumber\\
&\left.\left.+\left(4m^2(u-m_h^2)^2-M^4t\right)\left(2(u-m_h^2)C_{ac}+(u-m_h^2)^2D_{abc}\right)\right)
\right.\nonumber\\
&\left.+\frac{M^4+4m^2s}{2m^2}\left(\frac{s}{tu-m_h^4}((s-2m_h^2)m^2C_{ab}+(s-4m_h^2)C_{cd})\right.\right.\nonumber\\
&\left.\left.-(t-2m_h^2) D_{bac}-(u-2m^2)D_{abc}+2m^2D_{acb}-2C_{cd}\right)\frac{}{}\hspace{-2mm}\right)\nonumber
\label{eq:vectorHiggsPair2}
\end{align}
The calculations and scanning are done by Madgraph 5~\cite{Alwall:2011uj}, where we have implemented these amplitudes in addition to that of the top quark and colored scalar. The results from scanning over the masses of the vector and scalars are shown in the first three graphs in Fig.~\ref{figure1} and \ref{figure2}. The values for the portal couplings $\lambda_v$ and $\lambda_s$ are chosen to be $(\,1.5,\,1.5)$, $(\,1.5,\,0.5)$ and $(\,0.5,\,1.0)$  respectively for these scans. As for the scanning over the portal couplings we chose the masses of the vector and scalars to be $m_V=700$~GeV and $m_S=300$~GeV respectively. The corresponding results are shown in the fourth graphs of Fig.~\ref{figure1} and \ref{figure2}. The scans are done for $\sqrt{s}=8$~TeV (13~TeV) run and corresponding results are shown by black solid ( dashed) lines with their enhancements compared to the SM. We also superimposed the bounds from the latest $h \to ZZ^*$ for CMS (ATLAS) result which is shown in solid ( dashed) blue lines. The red line indicates the points which give the same results as the SM.

For $h\to \gamma\gamma$ the dominant contribution comes from the $W$ loop whose loop function is roughly 20 times larger than that of the scalar. This large loop function makes the octet vector contribution dominant for comparable masses and couplings. Also it has opposite sign compared to the scalar. For most of the plots we have two allowed regions. In the graphs from the mass scan, the upper allowed regions belong to the case where the contributions from the vector and scalars roughly cancel out while the lower regions belong to the case where large negative contribution from lighter vector octet completely dominates over others and the total result happens to end up being equal to the SM case. The latter case also appears as the upper allowed region for the graphs from the coupling scan and vice versa. The generic feature of the graphs can be easily understood in the following manner. For the low mass of the vector we sea roughly parallel curves (this is especially clear on the second plot of Fig.~\ref{figure1}). Since these are the cases for the vector particle domination the results hardly depend on the scalar mass. As for the higher masses of the vector and lower masses of the scalar we also see very small dependency on the vector mass as it should. As for the coupling scans the particular choice we made for the masses  we see the enhancements are essentially determined by the value of the vector mass.
	
	In both cases we see sizable enhancement in the Higgs pair productions in some parts of the parameter space that are still allowed by the latest $h\to ZZ^*$ measurements. The enhancements are especially pronounced for the complex vectors and scalars due to the larger Dynken index. As these particles are searched at the LHC a possibility of the scenario presented here will be tested or ruled out.


\section{Conclusion}

In this talk we present a partial result of ongoing project where we study the Higgs pair production due to color octet particles with spin zero and one. The cross-section compared the SM is calculated scanning over the masses of these particles as well as their portal couplings to the Higgs doublet for fixed masses. The change in the rate can be an order of magnitude or even larger compared to the SM case.

\begin{acknowledgments}
The author is thankful to the organizers of Toyama International Conference on HPNP 2015 for the opportunity to present his results. The funding from Ministry of Education, Culture and Science of Mongolia under the Innovation Project Grant for Postdoctoral Research is acknowledged. 
\end{acknowledgments}

\bigskip 

\begin{thebibliography}{99} 
\bibitem{atlas:2012gk}
  G.~Aad {\it et al.}  [ATLAS Collaboration],
  Phys.\ Lett.\ B {\bf 716} (2012) 1,
  S.~Chatrchyan {\it et al.}  [CMS Collaboration],
  Phys.\ Lett.\ B {\bf 716} (2012) 30.
\bibitem{Eboli:1987dy} 
  O.~J.~P.~Eboli, G.~C.~Marques, S.~F.~Novaes and A.~A.~Natale,
  Phys.\ Lett.\ B {\bf 197}, 269 (1987).
%
  E.~W.~N.~Glover and J.~J.~van der Bij,
  Nucl.\ Phys.\ B {\bf 309}, 282 (1988),
%
  D.~A.~Dicus, C.~Kao and S.~S.~D.~Willenbrock,
  Phys.\ Lett.\ B {\bf 203}, 457 (1988).
\bibitem{Jikia:1992mt} 
  G.~V.~Jikia,
  Nucl.\ Phys.\ B {\bf 412}, 57 (1994).
\bibitem{Barger:2013jfa} 
  See for example V.~Barger, L.~L.~Everett, C.~B.~Jackson and G.~Shaughnessy,
  %
\bibitem{Belyaev:1999mx} 
 %
  T.~Plehn, M.~Spira and P.~M.~Zerwas,
  Nucl.\ Phys.\ B {\bf 479}, 46 (1996)
  [Erratum-ibid.\ B {\bf 531}, 655 (1998)],
  A.~Belyaev~{\it et al.}, 
  Phys.\ Rev.\ D {\bf 60}, 075008 (1999),
%
  E.~Asakawa~{\it et al.},~
  Phys.\ Rev.\ D {\bf 82}, 115002 (2010),
 %
  G.~D.~Kribs and A.~Martin,~
  Phys.\ Rev.\ D {\bf 86}, 095023 (2012),
  N.~Haba, et al,
  Phys.\ Rev.\ D {\bf 89}, no. 1, 015018 (2014);
 T.~Enkhbat,
  JHEP {\bf 1401}, 158 (2014).
\bibitem{Alwall:2011uj}
  J.~Alwall, M.~Herquet, F.~Maltoni, O.~Mattelaer and T.~Stelzer,
  JHEP {\bf 1106} (2011) 128.

\end{thebibliography}

\end{document}